\documentclass[fleqn,usenatbib]{mnras}

\usepackage[T1]{fontenc}
\usepackage{ae,aecompl}
\usepackage{datatool}

\usepackage{graphicx}	% Including figure files
\usepackage{amsmath}	% Advanced maths commands
\usepackage{amssymb}	% Extra maths symbols
\newcommand\textlcsc[1]{\textsc{\MakeLowercase{#1}}}
\usepackage{newtxtext,newtxmath}
\title[A highly-resolved simulation of the Milky Way]{Determining the full satellite population of a Milky Way-mass halo in a highly resolved cosmological hydrodynamic simulation}

\author[R. J. J. Grand et al.]{\parbox[t]{\textwidth}{
Robert J. J. Grand$^{1,2,3}$\thanks{E-mail: rgrand@iac.es}, Federico Marinacci$^{4}$, R{\"u}diger Pakmor$^{1}$, Christine M. Simpson$^{5,6}$, Ashley J. Kelly$^{7}$, Facundo A. G{\'o}mez$^{8,9}$, Adrian Jenkins$^{7}$, Volker Springel$^{1}$, Carlos S. Frenk$^{7}$, Simon D. M. White$^{1}$ }
\vspace{10pt}
\\ 
$^1$Max-Planck-Institut f\"{u}r Astrophysik, Karl-Schwarzschild-Str. 1, 85748 Garching, Germany\\
$^2$Instituto de Astrof\'isica de Canarias, Calle Vía L\'actea s/n, E-38205 La Laguna, Tenerife, Spain\\
$^3$Departamento de Astrof\'isica, Universidad de La Laguna, Av. del Astrof\'isico Francisco S\'anchez s/n, E-38206, La Laguna, Tenerife, Spain\\
$^4$Department of Physics \& Astronomy ``Augusto Righi'', University of Bologna, via Gobetti 93/2, 40129 Bologna, Italy\\
$^5$Department of Astronomy \& Astrophysics, The University of Chicago, Chicago, IL 60637, USA\\
$^6$Enrico Fermi Institute, The University of Chicago, Chicago, IL  60637, USA\\
$^7$Institute for Computational Cosmology, Department of Physics, Durham University, South Road, Durham, DH1 3LE, UK\\
$^8$Instituto de Investigaci\'on Multidisciplinar en Ciencia y Tecnolog\'ia, Universidad de La Serena, Ra\'ul Bitr\'an 1305, La Serena, Chile\\
$^9$Departamento de Astronom\'ia, Universidad de La Serena, Av. Juan Cisternas 1200 Norte, La Serena, Chile}

\date{Accepted XXX. Received YYY; in original form ZZZ}

\pubyear{2021}

\begin{document}
\label{firstpage}
\pagerange{\pageref{firstpage}--\pageref{lastpage}}
\maketitle

\begin{abstract}
We investigate the formation of the satellite galaxy population of a Milky Way-mass halo in a very highly resolved magneto-hydrodynamic cosmological zoom-in simulation (baryonic mass resolution $m_b =$ 800 $\rm M_{\odot}$). We show that the properties of the central star-forming galaxy, such as the radial stellar surface density profile and star formation history, are: i) robust to stochastic variations associated with the so-called ``Butterfly Effect''; and ii) well converged over 3.5 orders of magnitude in mass resolution. We find that there are approximately five times as many satellite galaxies at this high resolution compared to a standard ($m_b\sim 10^{4-5}\, \rm M_{\odot}$) resolution simulation of the same system. This is primarily because 2/3rds of the high resolution satellites do not form at standard resolution. A smaller fraction (1/6th) of the satellites present at high resolution form and disrupt at standard resolution; these objects are preferentially low-mass satellites on intermediate- to low-eccentricity orbits with impact parameters $\lesssim 30$ kpc. As a result, the radial distribution of satellites becomes substantially more centrally concentrated at higher resolution, in better agreement with recent observations of satellites around Milky Way-mass haloes. Finally, we show that our galaxy formation model successfully forms ultra-faint galaxies and reproduces the stellar velocity dispersion, half-light radii, and $V$-band luminosities of observed Milky Way and Local Group dwarf galaxies across 6 orders of magnitude in luminosity ($10^3$-$10^{9}$ $\rm L_{\odot}$). 
\end{abstract}


%%%%%%%%%%%%%%%%%%%%%%%%%%%%%%%%%%%%%%%%%%%%%%%%%%

%%%%%%%%%%%%%%%%% BODY OF PAPER %%%%%%%%%%%%%%%%%%

\section{Introduction}

In the Lambda Cold Dark Matter ($\Lambda$CDM) cosmological framework, Milky Way-mass haloes assemble hierarchically, accreting many smaller dark matter haloes which then become subhaloes of the main system \citep{FWE85}. As a consequence, this theoretical model predicts that galaxies should be surrounded by populations of lower-mass galaxies. The abundance, properties, and distribution of these satellite galaxies depend on a wide range of galaxy formation physics (such as star formation and feedback) and the merging and disruption of subhaloes as they come into proximity with the central galaxy. As such, satellite galaxies provide critical tests of both the $\Lambda$CDM paradigm and galaxy formation models on small scales \citep[e.g.][]{DBM14,SFF16,BB17,MFB17,RJ18,WHP19,APR20,GPN21}.

The high degree of complexity and non-linearity involved necessitates the use of numerical simulations. For simulations in which baryonic physics is neglected (collisionless dark matter-only -- DMO -- simulations), the formation of haloes like that of the Milky Way has been studied in exquisite resolution using the ‘zoom-in’ technique \citep[e.g. the Via Lactea and Aquarius simulations; see][respectively]{DKM07,SWV08}. However, modelling the formation of large spiral disc galaxies like the Milky Way requires the inclusion of baryonic  processes within the $\Lambda$CDM cosmological environment, and has proven to be an exceedingly challenging endeavour \citep{NO17} because it involves many complex and interlinked astrophysical processes spanning a large dynamic range. Early attempts resulted in bulge-dominated systems - a consequence of the well-known overcooling problem and angular momentum catastrophe \citep{KG91,ANS03}. Nowadays, however, many groups have overcome these problems through  the inclusion of strong stellar feedback processes \citep[see][and references therein]{VMT20}. Several recent cosmological (magneto-)hydrodynamical simulations have produced disc-dominated, star-forming galaxies that match a number of key observables, such as rotation curves, stellar and cold gas structure, and present-day star formation rates \citep[e.g.][]{MPS14,WDS15,FMP20,LCG20,ARF21}. Such realistic central galaxies then serve as a platform to study the formation and properties of their satellite galaxies \citep{Sawala_15,SFF16,Wetzel+Hopkins+Kim+16,SGG17,BMD19,SWT20,SSNR20,ABC21}. In particular, the recent work of \citet{ABC21} has shown that it is becoming possible to  make predictions for the properties of ultra-faint satellite galaxies in simulations of Milky Way-mass haloes. This new regime offers interesting tests of our galaxy formation models, particularly with the recent discovery of dozens of ultra-faint dwarfs in and around the Milky Way \citep{Simon19}.

The physical drivers that shape the observed populations of satellite galaxies can be split into two halves: first, the formation of luminous galaxies and which dark matter haloes they occupy \citep{Benson_00,BWB03}; and second, their subsequent dynamical evolution and possible disruption. In the last decade, there has been growing interest in how tidal disruption shapes the present day satellite population, with a particular focus on the role of the baryonic disc potential relative to dark matter only simulations \citep{DSH10,PBW10,ZBW12,BZ14,Wetzel+Hopkins+Kim+16,Sawala_17,GWB17,Richings_20}. Recent studies have questioned the effectiveness of tidal disruption, with some arguing that it is not needed to solve the ``missing satellites'' problem \citep{KPH18,FMN19} and that too effective tidal disruption may in fact invert the problem \citep{KBG19}. Other, idealised dark matter-only $N$-body simulations claim that satellites orbiting in close proximity to the central galaxy experience some degree of ``artificial tidal disruption'' \citep{vdBO18,EP20,EN21,GBJ21}, with effects worsening for faint galaxies typically represented by a fewer number of particles relative to bright galaxies.

Tidal disruption effects manifest themselves strongly in the radial distribution of satellites: if it is too effective (artificial or otherwise), satellites are depleted preferentially near the central galaxy such that the radial distribution of galaxies is not as concentrated as suggested by observations. Indeed, precisely this discrepancy between numerical simulations and observations has been noted by several studies \citep[e.g.][]{BDB20,CGP20}, and numerical resolution has become the focal point of  attention. 

Recent semi-analytic work \citep{NCJ18,LGW19,BDB20} has demonstrated the importance of high resolution in simulations to help solve the problem, and thus explain the abundance and radial distribution of the full satellite populations of Milky Way-mass haloes. These models implement analytic prescriptions to treat the dynamical evolution (such as dynamical friction and tidal mass loss) of ``orhpan'' galaxies - galaxies whose dark matter haloes fall below the resolution limit - to approximate a higher resolution far beyond what can be currently achieved by the highest resolution hydrodynamical cosmological zoom-in simulations. However, predictions from the latter are crucial to better understand how the joint effects of tidal disruption and (perhaps more importantly) the {\it formation} of faint galaxies shape satellite galaxy populations in Milky Way-mass haloes.

In this study, we make a significant step toward addressing these issues by performing a magneto-hydrodynamical simulation for the formation of a Milky Way-mass galaxy that has a baryonic mass-resolution of $\sim 800$ $\rm M_{\odot}$. This simulation was performed with the Auriga galaxy formation model \citep{GGM17}, which we describe in Sec.~\ref{sec2}. In Sec.~\ref{sec3}, we present a resolution study of this galaxy across 3.5 orders of magnitude in particle mass, and demonstrate that the central galaxy shows a good level of convergence in the stellar surface density radial profile and the star formation history. We investigate the subhalo and satellite population of the main halo, and show that while subhaloes matched at different resolutions show good convergence, the subhalo/stellar mass functions highlight that twice as many satellites are present at the final time in our highest resolution compared to our standard resolution simulation (a factor of 64 in mass resolution). We investigate how the formation and survival of satellites depends on resolution and on their orbital properties at infall, and how artificial disruption modifies the radial distribution of satellites around their host haloes. Finally, we show that the satellite population of our highest resolution simulation includes analogues of observed ultra-faint galaxies, and reproduces several observable scaling relations of faint dwarf galaxies in the Local Group. In Sec.~\ref{Discussion}, we discuss our results in the context of recent simulation efforts and observational findings. In Sec.~\ref{Conclusions}, we present our conclusions and offer additional applications for our high resolution simulation for several aspects of galaxy evolution.

\section{Methodology}
\label{sec2}

\subsection{Simulations}

The simulated galaxy presented in this paper is a re-simulation of one of the Milky Way-mass systems from the \textlcsc{Auriga} project, \citep[][]{GGM17,GHF18}. These were specifically selected to be between $1$-$2\times 10^{12} \rm M_{\odot}$ in total mass ($M_{200}$), which we define as the mass contained inside the radius at which the mean enclosed mass volume density equals 200 times the critical density for closure. They were initially selected from the $z=0$ snapshot of a parent dark matter only cosmological simulation of comoving periodic box size 100 Mpc, with the standard $\Lambda$CDM cosmology. The adopted cosmological parameters are $\Omega _m = 0.307$, $\Omega _b = 0.048$, $\Omega _{\Lambda} = 0.693$ and a Hubble constant of $H_0 = 100 h$ km s$^{-1}$ Mpc$^{-1}$, where $h = 0.6777$, taken from \citet{PC13}. At $z=127$, the resolution of the dark matter particles of this halo is increased and gas is added to create the initial conditions of the zoom. 

The simulations were run with the magneto-hydrodynamic code \textlcsc{AREPO} \citep{Sp10}, and a model that includes many astrophysical processes deemed important for galaxy formation: primordial and metal line cooling, a model for star formation that activates for gas densities larger than $0.1$ atoms $\rm cm^{-3}$ \citep{SH03}, magnetic fields \citep{PMS14,PGG17,PGP18}, gas accretion onto black holes and energetic feedback from AGN and supernovae type II (SNII) \citep[see][for more details]{VGS13,MPS14,GGM17}. Each star particle is treated as a single stellar population of given mass, age and metallicity. Stellar mass loss and metal enrichment from type Ia supernovae (SNIa) and Asymptotic Giant Branch (AGB) stars are modelled according to a delay time distribution. We track a total of 9 elements produced via these stellar evolutionary processes: H, He, C, O, N, Ne, Mg, Si and Fe, in addition to 6 $r$-process elements produced by neutron star mergers \citep{vdVPG20}. We have modified the implementation of the uniform UV background from previous Auriga simulations in which the UV background was switched from 0 to full strength at $z=6$. In all simulations presented here, the background field is gradually increased such that it reaches full strength at redshift 6\footnote{This change produces small quantitative differences in the faint end of the satellite luminosity function. However, it does not make an overall qualitative difference in the properties of the satellite populations or host galaxies.}.

The \textlcsc{AURIGA} model has been shown to produce realistic spiral disc galaxies that are broadly consistent with a number of observations including star formation histories, stellar masses, sizes and rotation curves of Milky Way-mass galaxies \citep{GGM17}, the distribution of HI gas \citep{MGP16}, the stellar halo properties of local galaxies \citep{MGG19}, stellar disc warps \citep{GWG16}, the properties and abundance of galactic bars \citep{Fragkoudi+Grand+Pakmor+19} and bulges \citep{GMG19}, and the properties of magnetic fields in nearby disc galaxies \citep{PGG17,PGP18}. 

In this study, we present a simulated galaxy with a mass resolution of $\sim 800$ $\rm M_{\odot}$ per baryonic element, and $ 6 \times 10^3$ $\rm M_{\odot}$ per dark matter particle, with a softening length of $90$ pc after $z=1$. We designate this resolution as ``Level 2'', in keeping with previous nomenclature for the Auriga simulations. The simulation took approximately 15 million CPU hours to complete. In the following, we compare the properties of the central galaxy and its satellite population in this simulation to lower resolution simulations of the same system, the specifications for which we list in Table 1.

\subsection{Matching subhaloes between resolution levels}
\label{match}

Our simulations were run with the on-the-fly SUBFIND halo finder \citep{SWT01} which initially identifies haloes with a Friends-of-Friends algorithm \citep{DEF85} and then separates them into disjoint gravitationally self-bound subhaloes (containing at least 20 particles). The simulations were postprocessed with the LHaloTree algorithm that tracks the progenitors and descendents of (sub)haloes at each simulation output time \citep{SW05}, and thus constructs a merger tree for each simulation. We use the merger tree information to track the orbital histories of all objects only after $z=4$, because of uncertainties in the identification of progenitors at earlier times \citep[see also][]{FDF20}. Of these objects, we first identify all luminous subhaloes that are either found within the virial radius of the central galaxy at $z=0$ or have merged some time in the past. 

For each object, we track the orbit of the main progenitor branch backwards and define the infall time as the time at which the object first crossed the virial radius into the main halo. We record the orbital coordinates at each snapshot prior to infall for each of the identified objects in each simulation; interactions with the host introduce unnecessary complications for matching orbital trajectories. We define the following simple metric to estimate the level of agreement between the coordinates of the $i$-th lower-resolution object and the $j$-th higher resolution object:

\begin{equation}
M_{i,j} = \frac{1}{N} \sum _k^{N} \sum _l^3 | x_{i,l,k} - x_{j,l,k}|
\end{equation}
where $x_{j,k,l}$ is the $l$-th coordinate of the $j$-th object in the $k$-th snapshot pair. We determine the earliest infall time of each considered $i$-$j$ pair, and sum over $N$ snapshots in the lower resolution simulation prior to this time - each lower resolution snapshot is matched to the nearest (in time) higher resolution snapshot. This ensures that each $M_{i,j}$ value is calculated using halo coordinates outside of the virial radius of the main halo and thus minimises orbital deviations that may reflect interactions within the main halo. We determine the matched higher-resolution halo for the $i$-th lower resolution halo by finding the minimum value of $M_{i,j}$ along the corresponding axis. In practice, we find that each matched pair has an orbit averaged metric of $\lesssim10$ kpc, which yields an excellent level of agreement between the orbits and masses of each standard resolution satellite/progenitor and its matched high-resolution counterpart prior to infall.

\begin{table}
\centering
 \begin{tabular}{|c|c|c|c|}
 \hline
Level & $m_{\rm DM}$ & $m_{\rm b}$ & $h_{\rm b}$ \\
 & ($\rm M_{\odot}$) & ($\rm M_{\odot}$) & (pc) \\
 \hline
2 & 4,600 & 850 & 94 \\
3 & $3.6\times10^4$ & 6,700 & 188 \\
4 & $2.9\times10^5$ & $5.4\times10^4$ & 375 \\
5 & $2.4\times10^6$ & $4.4\times10^5$ & 750 \\
6 & $2.0\times10^7$ & $4.2\times10^6$ & 1,500 \\
 \hline
\end{tabular}
\caption{Summary of the numerical resolution for each simulation analysed in this paper. The columns are: resolution ``level'' of the run; dark matter particle mass; baryonic target particle/cell mass; softening length.}
\label{tabRes}
\end{table}

\begin{figure}
\centering
\includegraphics[scale=2.4,trim={0 0 0 0}, clip]{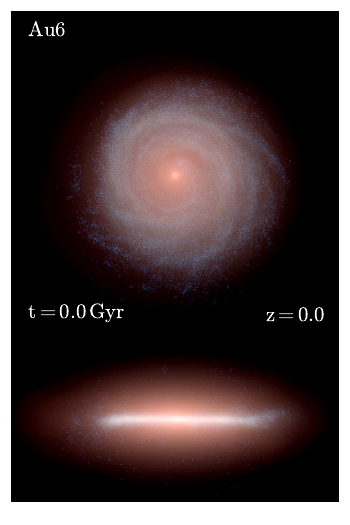}
\caption{Face-on (upper) and edge-on (lower) stellar light projection of the $K$-, $B$- and $U$-band luminosity of stars in our high-resolution simulated galaxy at present day. Bluer (redder) colours indicate younger (older) stars. The $x$-axis is 50 kpc in length.}
\label{fig1}
\end{figure}

\begin{table*}
\centering
 \begin{tabular}{|c|c|c|c|c|c|c|c|c|c|c|}
 \hline
Level & $M_{200}$  & $M_{*}$  & $M_{\rm cold \, gas}$ & $M_{\rm disc}$ & $R_{\rm exp}$ & $M_{\rm bulge}$ & $r_{\rm eff}$ & D/(D+B) & $\tau_{*}$ & $Z_*$\\
 & ($10^{12}{\rm M}_{\odot}$) & ($10^{10}{\rm M}_{\odot}$) & ($10^{10}{\rm M}_{\odot}$) & ($10^{10}{\rm M}_{\odot}$) & (kpc) & ($10^{10}{\rm M}_{\odot}$) & (kpc) & & (Gyr) &  \\
 \hline
2 & 1.02 & 6.72 & 0.79 & 5.67 & 3.62 & 0.31 & 0.87 & 0.95 & 7.42 & 0.030 \\
3 & 1.04 & 6.89 & 0.85 & 5.39 & 4.03 & 1.18 & 1.27 & 0.82 & 6.58 & 0.032 \\
4 & 1.03 & 5.15 & 0.72 & 4.55 & 4.85 & 0.60 & 1.09 & 0.88 & 6.70 & 0.027 \\
5 & 1.03 & 3.84 & 0.30 & 3.51 & 4.82 & 0.33 & 1.19 & 0.91 & 6.97 & 0.023 \\
6 & 0.99 & 3.36 & 0.73 & 2.81 & 3.28 & 0.07 & 0.83 & 0.97 & 6.42 & 0.021 \\
 \hline
\end{tabular}
\caption{Galaxy parameters for each simulation. The columns are: resolution level; halo mass; total stellar mass within one tenth of the virial radius; the cold ($T\lesssim10^4$ K) gas mass contained within 1 tenth of the virial radius; the stellar disc mass; exponential disc scale length; stellar bulge mass; bulge effective radius; ratio of the disc stellar mass ($D$) to the sum of the disc and bulge stellar mass ($D+B$); median stellar age and metallicity of all star particles within one tenth of the virial radius. The disc and bulge components have been derived from a simultaneous fit of a S{\'e}rsic and exponential profile to the stellar mass surface density \citep[as done in][]{GGM17}.}
\label{t2}
\end{table*}

\section{Results}
\label{sec3}

\begin{figure*}
\centering
\includegraphics[scale=0.95,trim={0 1.2cm 1.cm 0}, clip]{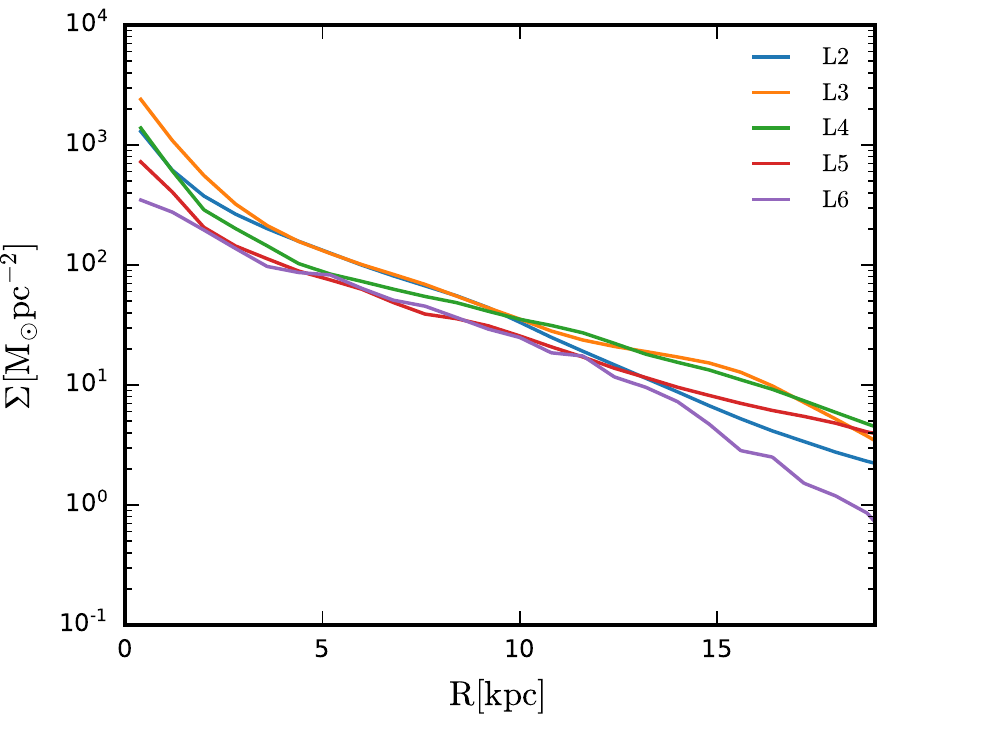}
\includegraphics[scale=0.95,trim={0 1.2cm 1.cm 0}, clip]{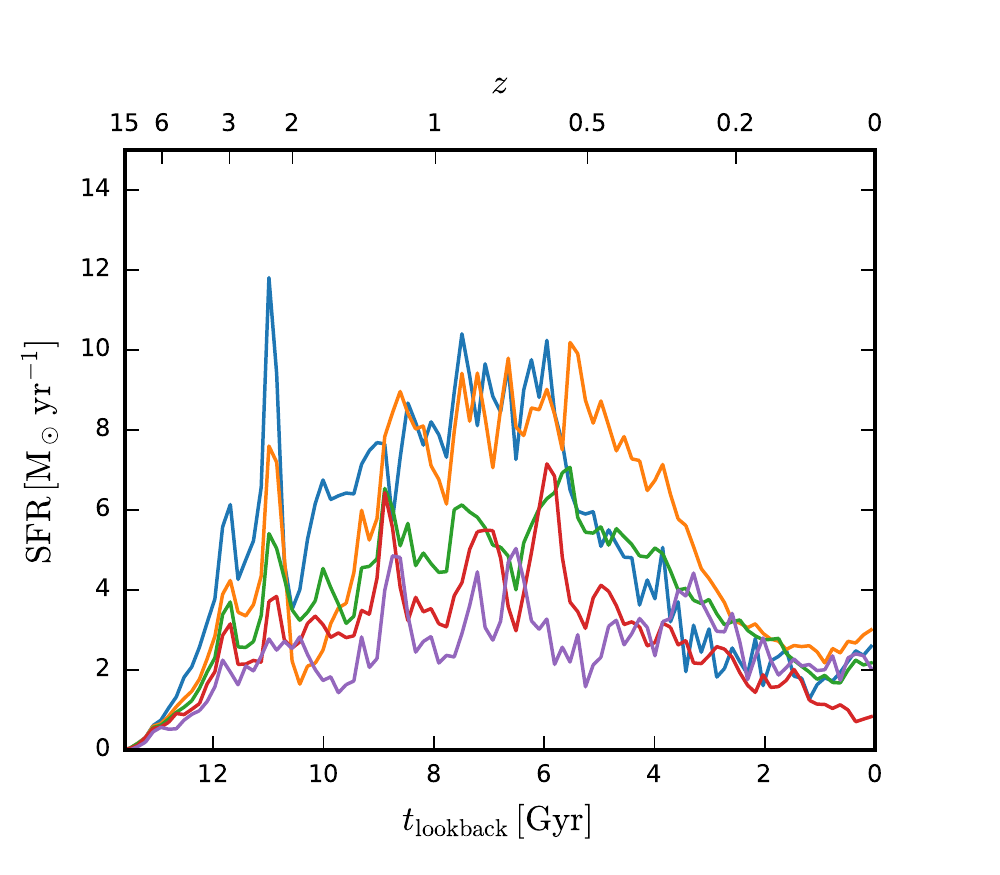}\\
\includegraphics[scale=0.95,trim={0 0 1.cm 0.2cm}, clip]{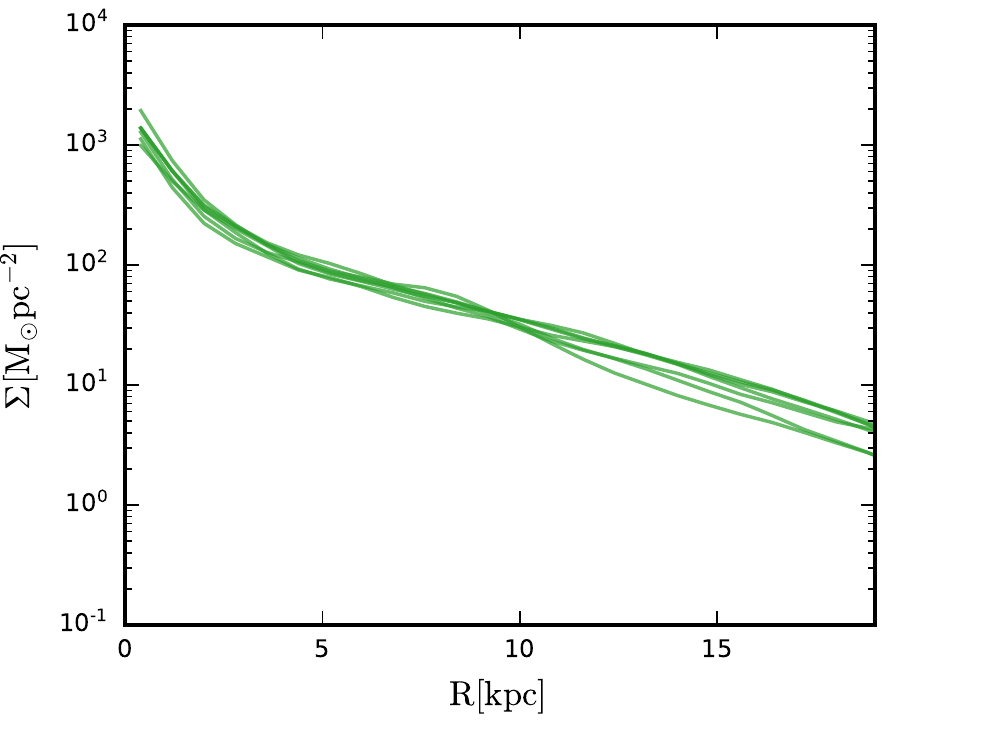}
\includegraphics[scale=0.95,trim={0 0 1.cm 1.4cm}, clip]{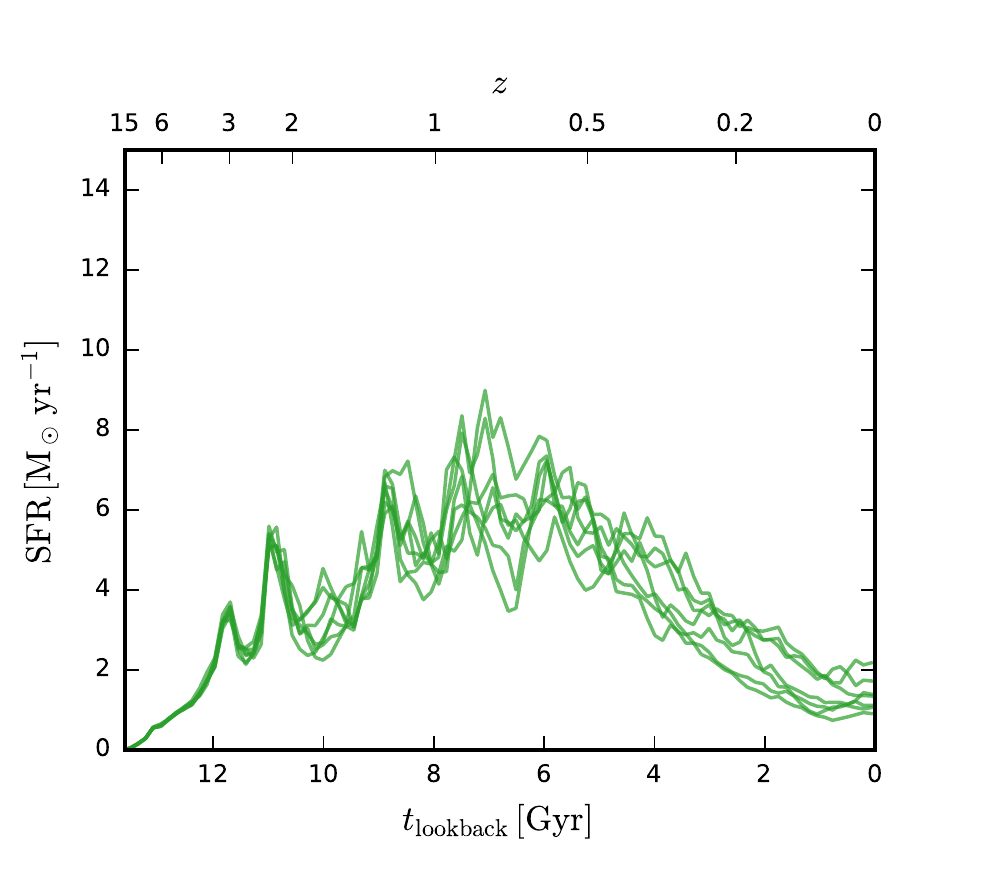}
\caption{Top left: the projected radial stellar surface density profile of our high-resolution simulated galaxy (denoted level 2) and of the same galaxy simulated at a series of lower mass resolution simulations that successively increase the particle mass by factors of 8. Top right: As the left panel, but for the star formation history (SFH). Bottom: as above, but for 7 different realisations of the standard resolution level 4 simulation; each simulation has exactly the same initial conditions but for the random seed, which is different in each case. The scatter in these curves thus reflects the stochastic variation associated with the ``butterfly effect'' \citep[e.g.][]{GBS19,KWW19}, which is small.}
\label{fig2}
\end{figure*}

\begin{figure*}
\includegraphics[width=\columnwidth,trim={0 1cm 0.3cm 0.7cm}, clip]{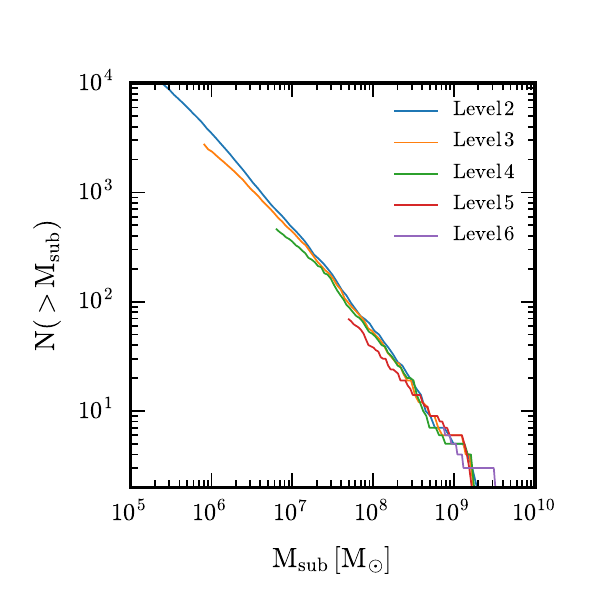}
\includegraphics[width=\columnwidth,trim={0 1cm 0.3cm 0.7cm}, clip]{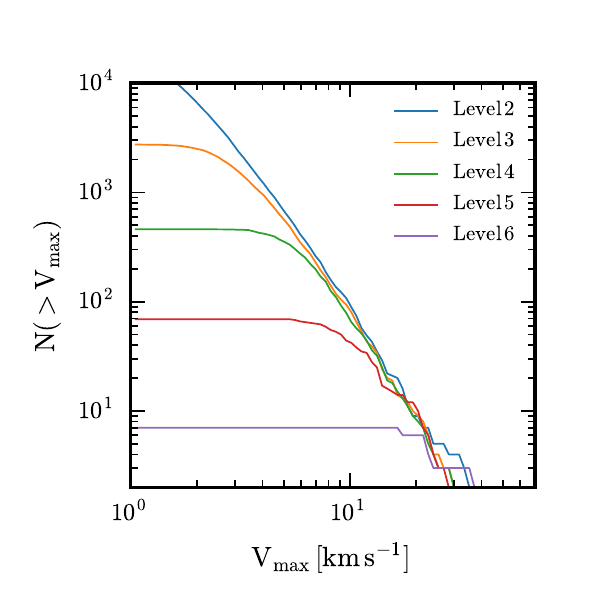}\\
\includegraphics[width=\columnwidth,trim={0 1.4cm 0.3cm 2.cm}, clip]{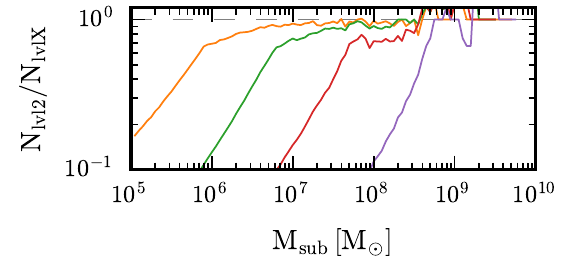}
\includegraphics[width=\columnwidth,trim={0 1.4cm 0.3cm 2.cm}, clip]{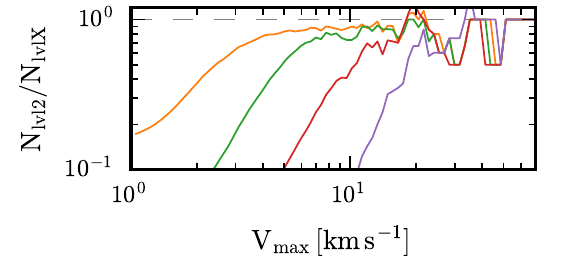}
\caption{Cumulative subhalo mass function (left) and subhalo $V_{\rm max}$ function (right) for all objects within 300 kpc of the centre of the main halo at the present day. The lower panels show the curves for each resolution normalised to that for the highest resolution (level 2) simulation. $V_{\rm max}$ has been corrected for the effects of softening following Eqn.~10 in \citet{SWV08}.}
\label{fig3}
\end{figure*}

\subsection{Global properties of the central galaxy}

Fig.~\ref{fig1} shows the face-on and edge-on stellar light projection of our simulated galaxy at redshift zero. This image is characterised by a blue, radially-extended star-forming galactic spiral disc, which surrounds a redder central stellar bulge. The edge-on view shows that the disc is slightly warped, and is embedded in a larger, redder stellar halo. The radial stellar surface density profile is presented in the top-left panel of Fig.~\ref{fig2}, alongside the profiles of the same object simulated at four lower resolutions, which together span a factor of 4096 in mass. We stress that the physical parameter settings for the galaxy formation model are unchanged at each resolution. The degree to which the surface density profiles vary over this notably large range in resolution is at most a factor of 2 for radii larger than about 3 kpc, with larger variations in the central kiloparsec. At most radii, the stellar surface density increases marginally with increasing resolution; however certain radii do not always reflect this trend, and there are signs that the 2 highest resolution simulations are more similar than the lower resolution simulations. 

These trends are borne out by the star formation histories shown in the top-right panel of Fig.~\ref{fig2}; the star formation rate increases marginally, particularly for lookback times greater than about 4 Gyr, with the notable exception that the level 2 and level 3 simulations are very similar earlier than approximately 6 Gyr lookback time, after which time the level 2 drops below the level 3 and becomes almost identical to the level 4 simulation. At lookback times $\lesssim4$ Gyr, all simulations become very similar and show no systematic trend in star formation rate. Note that the salient trends of the star formation histories: the peak at about redshift 1 and steady subsequent decline to star formation rates of order 1 $\rm M_{\odot} \, yr^{-1}$ at redshift zero, are retained across 3.5 orders of magnitude in resolution. We note that an unexpected problem associated with our black hole centering procedure (commonly adopted in cosmological simulations) revealed itself in our level 2 simulation: after $z\sim 1$; the black hole (mass $\sim 10^{7}M_\odot$) wanders around the disc at typical distances of order a few kiloparsecs from the galactic centre. This unexpected behaviour may negatively impact the convergence of the late-time properties of the central galaxy. Nevertheless, we find that the total stellar, disc, bulge, and gas masses, as well as the scale radii and mean stellar population properties (listed in Table~\ref{t2}) vary little across all resolution levels with the exception of level 6. We stress that this level of convergence has not previously been demonstrated in cosmological hydrodynamic simulations for such a large dynamical range in resolution.

\begin{figure*}
\includegraphics[width=\columnwidth,trim={0.3cm 0.3cm 0.3cm 0.5cm}, clip]{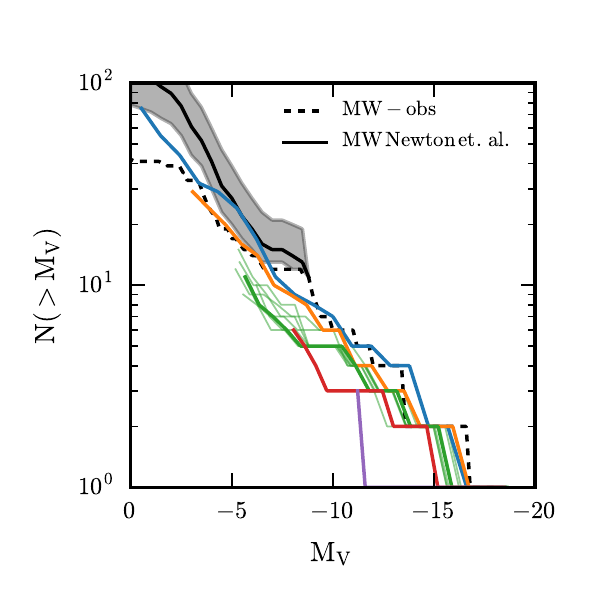}
\includegraphics[width=\columnwidth,trim={0.3cm 0.3cm 0.3cm 0.5cm}, clip]{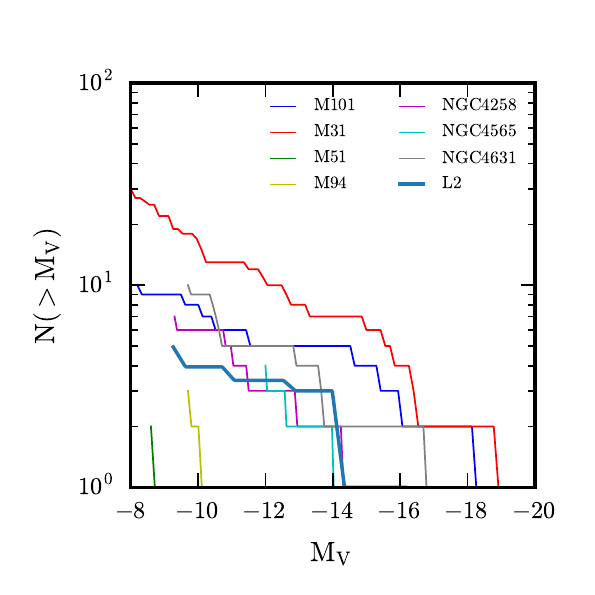}
\caption{Left: Cumulative $V$-band absolute magnitude function of all satellites within 300 kpc of the centre of the main halo at the present day, for all resolution levels (colour is as for other figures: L2: blue; L3: orange; L4: green; L5: red; L6: purple). We show also the distribution of 6 additional realisations of the L4 run with thin green lines to illustrate the scatter from the Butterfly effect. Observational data for classical dwarf galaxies, SDSS DR9, and the DES survey \citep[see Table A1 of][]{NCJ18} are shown by the dashed curve, and the median and 2$\sigma$ of the completeness-corrected data for $M_V>-8.8$ listed in Appendix E of \citep{NCJ18} are shown by the solid black and grey shaded region, respectively. Right: As the left panel, but for high resolution satellites with $\mu _{V,\rm eff} < 28.3$ and $M_V < -9$ within a projected radius of 150 kpc of the central galaxy averaged over 1000 lines of sight. This mimics the selection function of the Milky Way mass galaxies in the Local Volume described in Table 3 of \citet[][]{CGP21}, the luminosity functions of which are shown by the thin lines.}
\label{fig4}
\end{figure*}

\subsubsection{The Butterfly Effect}

At face value, resolution studies of individual systems like those presented in Fig.~\ref{fig2} assume that each realisation is a unique prediction for a given model and set of initial conditions. However, we know that this is not true in practice. In particular, recent studies have highlighted that running the same code on the same initial conditions with a different random seed can change the result substantially \citep[e.g.][]{GBS19,KWW19}. The extent of such variations depends on the details of the various components of a given galaxy formation model and must be understood in order to assess the predictive power of the model. 

The lower-left and lower-right panels of Fig.~\ref{fig2} show the radial stellar surface density profiles and star formation histories, respectively, for 7 realisations (each with a different random seed) of our level 4 setup. In general, the scatter in each of these plots is very small; the normalisations and slopes of the surface density profiles are similar at all radii and the shape of the star formation histories and their peak values differ by less than 2 $\rm M_{\odot} \, yr^{-1}$ at any given time. For each plot, the scatter is smaller than that of the resolution study in the upper panels of Fig.~\ref{fig2}. We have also verified that the satellite luminosity function shows a similarly low-level of scatter associated with the random seed. This reassures us that the Auriga model does not have a significant ``Butterfly Effect'' problem, and that most of the variations in our resolution study arise from changes in resolution as opposed to variations associated with numerical stochasticity.

\subsection{The Subhalo population}
\label{subpop}

\begin{figure}
\includegraphics[width=\columnwidth,trim={0.2cm 1.2cm 0.3cm 1.8cm}, clip]{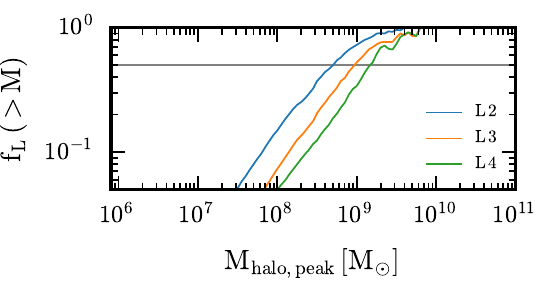}
\includegraphics[width=\columnwidth,trim={0.2cm 2.2cm 0.3cm 1.8cm}, clip]{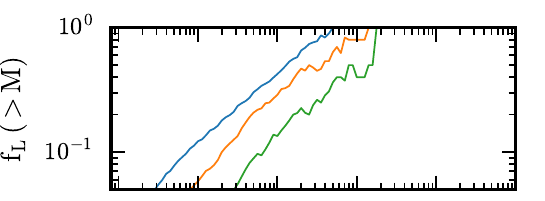}
\includegraphics[width=\columnwidth,trim={0.2cm 0 0.3cm 0.8cm}, clip]{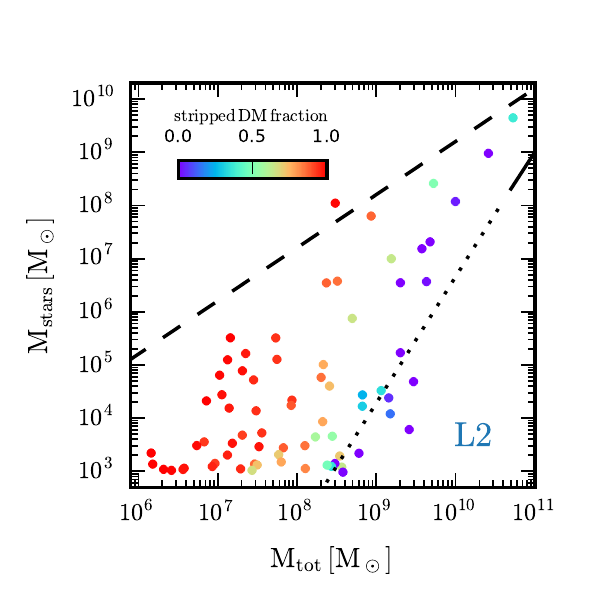}
\caption{Top: Fraction of subhaloes above a given peak halo mass that are luminous at the present day for all objects within 1 Mpc of the central galaxy in all simulations. Middle: As the top panel, but for the present day halo mass. Bottom: Stellar mass-halo mass relation for all galaxies within 1 Mpc of the central galaxies in our level 2 simulation. Points are colour-coded according to the fraction of dark matter a given object has lost since it reached its peak mass. The universal baryon fraction is indicated by the dashed black line, and the solid-dotted line is the abundance matching relation from \citet{MNW13}: the dotted portion highlights the mass regime where this relation is uncertain. }
\label{figmstar}
\end{figure}

In this section, we focus on the properties of the subhalo/satellite population and how they are affected by resolution. The left and right panels of Fig.~\ref{fig3} shows the cumulative subhalo mass and $V_{\rm max}$ functions for all objects within a radius of 300 kpc of the central galaxy for each of our 5 resolution levels. Each curve is characterised by a power law that rolls off for objects of lower and lower mass as resolution increases. The lower panels of this figure normalise these curves by that of the highest resolution simulation to more clearly highlight the differences between resolution levels. From these plots, we infer that $\gtrsim100$ dark matter particles per subhalo are required for lower-resolution simulations to reproduce $\gtrsim90\%$ of the subhaloes at $z=0$ in our highest resolution simulation. For our standard resolution simulations (level 4), this translates to a subhalo mass, $M_{\rm sub}\gtrsim 5\times 10^7 \, \rm M_{\odot}$, or equivalently, $V_{\rm max} \gtrsim 10 \, \rm km/,s^{-1}$. However, convergence drops to $\sim 70\%$ and $\sim 10\%$ for subhalo masses $M_{\rm sub}\gtrsim 10^7 \, \rm M_{\odot}$ and $M_{\rm sub}\gtrsim 10^6 \, \rm M_{\odot}$, respectively.

The left panel of Figure~\ref{fig4} shows the cumulative satellite luminosity function for all satellites within 300 kpc of the host halo centre in each of our simulations. The shape of the function is generally preserved at each resolution, with the function extending to fainter galaxies as resolution increases. The number of satellites fainter than $M_V\sim -10$ seems to increase with resolution -- we will discuss this further in the next section. The abundance of bright satellites, in contrast, appears to show little variation for most resolution levels, although there is an increment of 1 at $M_V \sim -13$ for our level 2 simulation relative to lower resolution simulations (with the exception of level 6, which is clearly too low a resolution to even capture the brightest satellites reliably). We observe a slight trend of increasing luminosity with increasing resolution for the brightest satellites, although this seems to be similar to the variation induced by the Butterfly effect; the scatter in different realisations of the level 4 luminosity function arguably accounts for much of this discrepancy. 

We compare the simulated satellite luminosity functions with recent observational data for the Milky Way's satellites analyzed by \citet{NCJ18} and with  the fitting function for the faint end of the luminosity function derived by the same authors. It is clear that our level 2 simulation produces the best fit to the observed luminosity function, and that this match becomes less good as resolution decreases. At the faint end, this is a good match to the data, although the extrapolation of the fit function from \citet{NCJ18} implies the abundance may still be underestimated for galaxies fainter than $M_V \gtrsim -4$.

Finally, we compare the level 2 luminosity function to those of several Milky Way-mass host haloes in the local volume in the right panel of Figure~\ref{fig4}. To make this comparison, we calculate the $V$-band surface brightness within the effective projected half-light radius, $\mu _{V,\rm eff}$, and absolute magnitude, $M_V$, of each satellite within 300 kpc of the main galaxy, and select only those with $M_V < -12$ and $\mu _{V,\rm eff} < 28.3$  \citep[as is also done in][for example]{FMB21}. We calculate the projected radius of these satellites along 1000 random lines of sight, and select those within a projected distance of 150 kpc of the central galaxy to mimic the apertures used in the observations. We divide the resulting luminosity function by 1000. The right panel of Figure~\ref{fig4} shows that this luminosity function lies within the scatter of the local volume observations.

The lower panel of Fig.~\ref{figmstar} shows the stellar mass as a function of total mass at $z=0$ for all level 2 galaxies (including satellite galaxies) in a 1 Mpc volume around the central galaxy. Each point is colour-coded to reflect the fraction of dark matter a given object has lost since it reached its peak mass. We clearly see that galaxies that have experienced more tidal stripping are found further to the left in this plot. This trend was first highlighted by \citet{Sawala_15} in the Apostle simulations \citep{Fattahi_16} and reported in \citet{SGG17} for the full sample of level 4 Auriga simulations. However, we  now see that it holds for stellar masses $< 10^6 \, \rm M_{\odot}$ and halo masses $10^6$ - $10^8 \, \rm M_{\odot}$. Notably, objects that have experienced the least dark matter mass loss lie very close to the extrapolated abundance matching curve \citep{MNW13}. Aside from the effects of tidal stripping, \citet{Sawala_15} showed that the abundance matching relation derived from higher masses cannot be extrapolated to the lower halo masses at which only a fraction of the haloes host a galaxy. These authors provide a correction to the abundance matching relation that takes this into account.

In the top and middle panels of Fig.~\ref{figmstar}, we plot the cumulative luminous fraction, $f_L$, as a function of the peak and $z=0$ halo mass, of all simulated haloes within a 1 Mpc sphere around the central galaxy. We observe a trend that, for a given value of $f_L$, the halo mass (both peak and $z=0$) above which $f_L$ of the subhalos are luminous decreases with increasing resolution. For example, at level 4, $f_L=0.5$ for $z=0$ masses $\sim 10^9 \, \rm M_{\odot}$, which drops to $\sim 10^8 \, \rm M_{\odot}$ at level 2. These panels may be compared with Figure~4 of \citet{MBA21} and Figure~2 of \citet{Sawala_16a}, respectively, who found a similar dependence on resolution. The level 2 Apostle simulations in the latter study, which have comparable resolution to our level 3 simulations, the 50\% occupation mass is $2\times 10^9 \, \rm M_{\odot}$, an order of magnitude larger than for our level 3 simulations. The main reason for this difference is the assumed redshift of reionization: 6 in our simulations but 11.5 in Apostle. The dependence of $f_L$ on the redshift of reionization was calculated by \citet{Benitez-Llambay_20} (see their Figure~11) who studied in detail how  the halo occupation fraction depends on the modelling of gas cooling, reionization  and star formation at high redshift. 

It is worth noting that \citet{NWB20} find, using a halo occupation model, that nearly all haloes with a peak halo mass greater than $\sim 3\times 10^8 \, \rm M_{\odot}$ contain a luminous galaxy. This is because such models employ analytic prescriptions to predict how galaxies and their host haloes evolve below the resolution limits of fully numerical simulations. This is consistent with the notion that the increase in luminous fraction for low mass haloes is a combination of the ability to form stars in small-mass haloes and the effects of tidal stripping, which we discuss below. 

\begin{figure}
\includegraphics[width=\columnwidth,trim={0.3cm 0 0.3cm 0.5cm}, clip]{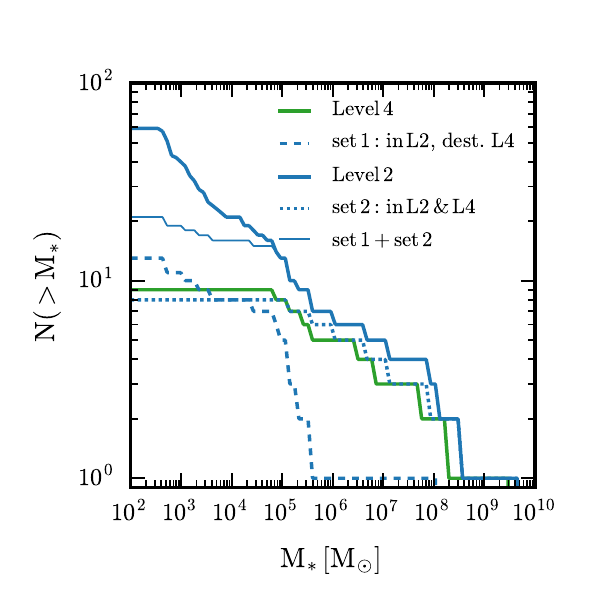}
\caption{Cumulative stellar mass function of satellites within the virial radius of the main halo for our standard level 4 (green) and high resolution level 2 (thick blue solid) simulations. These simulations are separated in baryon mass resolution by a factor of 64. The function for satellites that survive at high resolution but disrupt at standard resolution is shown by the dashed blue curve. High-resolution counterparts of the level 4 bright satellites (blue dotted line in Fig.~\ref{fig4}) have a very similar mass function to the level 2 satellites. The thin blue line is the sum of the dashed and dotted lines; the discrepancy between this line and the thick solid line reflects the objects that formed only in the high resolution simulation.}
\label{fig4c}
\end{figure}

\subsubsection{Satellite evolution and disruption}

\begin{figure*}
\includegraphics[scale=1.5,trim={0cm 0.9cm 0.3cm 0cm}, clip]{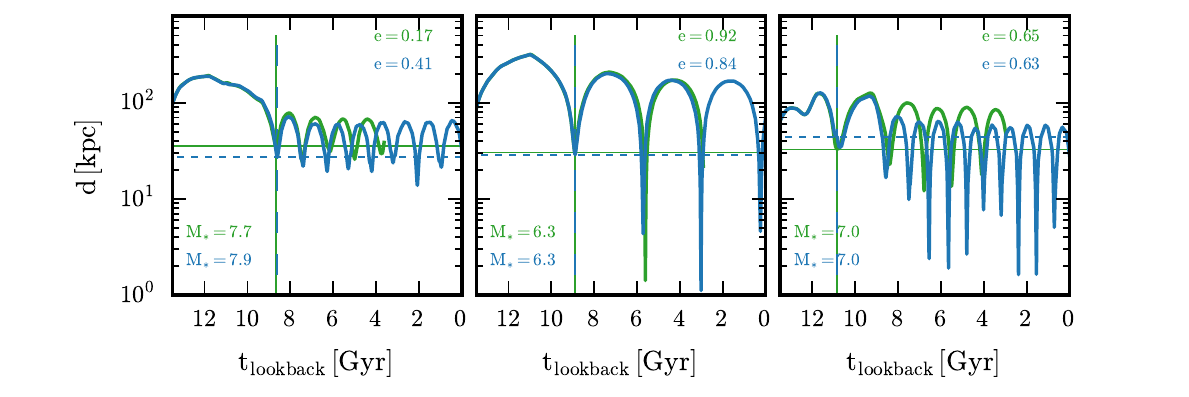}\\
\includegraphics[scale=1.5,trim={0.cm 0.9cm 0.3cm 0cm}, clip]{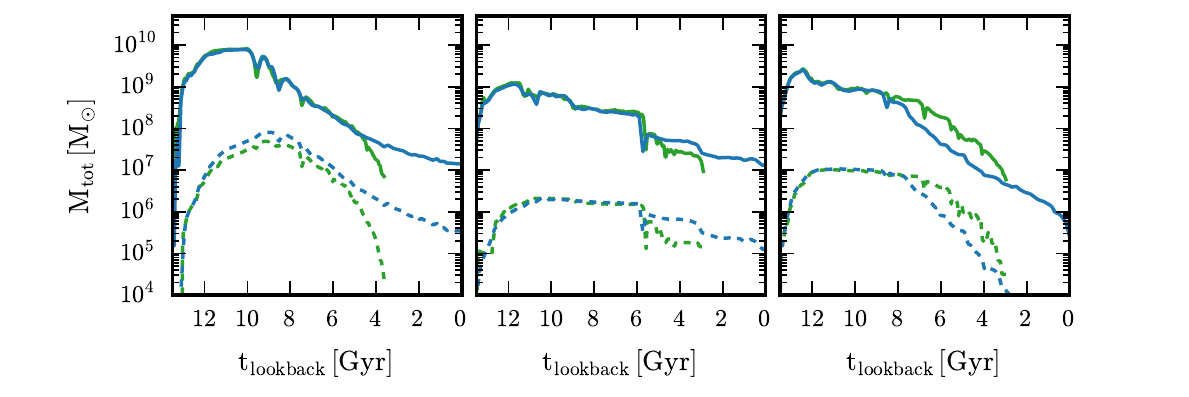}\\
\includegraphics[scale=1.5,trim={0.cm 0cm 0.3cm 0cm}, clip]{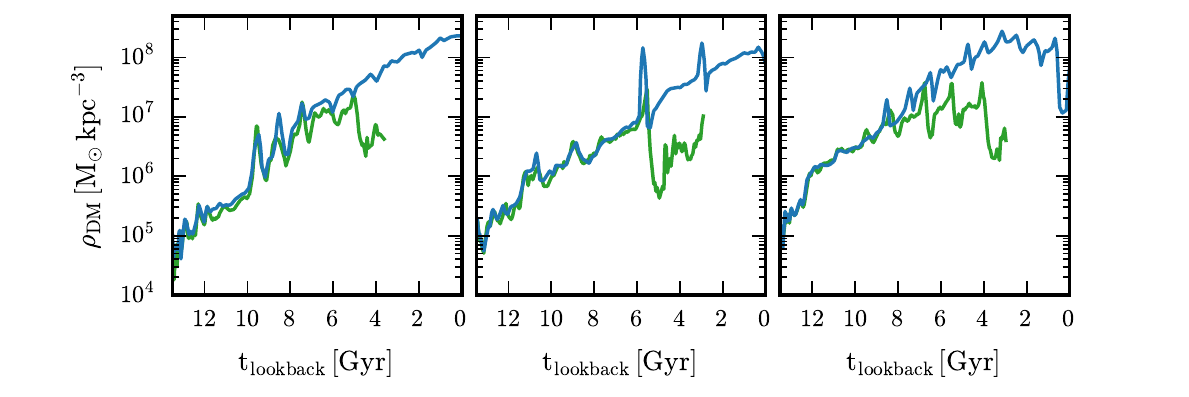}\\
\caption{Evolution of three luminous subhaloes of varying orbital eccentricities at infall (columns) in the standard level 4 (green)  and the high-resolution (level 2) simulations. The rows show evolution of galactocentric distance (top), the total and stellar masses (middle; solid and dashed lines, respectively), and the dark matter density, defined as the mass enclosed within the half-mass radius divided by its volume (bottom). The impact parameter and infall time, defined as the distance and time of the first pericentric passage after infall, are indicated by the horizontal and vertical lines, respectively. In each case, the satellite survives at high resolution, but disrupts at standard resolution. All measured quantities show very similar evolution before infall (except for the stellar mass of the satellite in the first column), but deviations occur with increasing pericentric passages. In particular, the dark matter density drops suddenly after sufficient mass loss has occurred at standard resolution, whereas it steadily increases at high resolution. Note that in the rightmost column, the pericentre after infall is marginally smaller at high resolution, which causes mass loss to occur more promptly compared to its standard resolution counterpart.}
\label{fig5}
\end{figure*}

In this section, we study the resolution dependence of subhalo formation and disruption and their relation to the abundance of satellites. For clarity, we focus on comparisons between the highest resolution simulation (level 2) and the ``standard'' resolution simulation (level 4), which has a mass resolution 64 times poorer. Comparisons involving the standard resolution simulation are interesting because they are of similar resolution to the highest resolution cosmological box simulations \citep[e.g. Illustris TNG;][]{PNS19}. 

Fig.~\ref{fig4c} shows the satellite stellar mass function of the level 2 (blue solid line) and the standard resolution level 4 (green solid) simulations. Two possible explanations for the offset in these two functions are: i) the stellar masses of individual  satellites are larger at higher resolution compared to lower resolution for the same objects; and ii) there are more objects of a given stellar mass (particularly near the low-mass end) at high resolution compared to lower resolution. To understand the relative importance of each, we first consider the impact of tidal disruption on matched objects in the two simulations. We use the method described in Sec.~\ref{match} to identify and match all surviving satellites and disrupted systems in both the standard and high-resolution simulations. In practice, we find a good match at level 2 for all the luminous satellites and progenitors identified after $z=4$ in the level 4 simulation.

The cumulative stellar mass function of all objects that survive in both the level 4 and level 2 resolution simulations (set 2; blue dotted) is very similar to the $z=0$ cumulative stellar mass function of the level 4 simulation. Interestingly, one of the surviving level 4 satellites is disrupted in the level 2 simulation, because in this case their orbital trajectories deviate strongly after infall; the level 2 subhalo inspirals whereas the level 4 counterpart has a large apocentre. The most massive level 2 satellites can be a factor $\sim 2$ larger than their level 4 counterparts; otherwise the small differences between these two distributions are similar to  stochastic variations induced by changes in random seed, as discussed in the preceding section. Figure~\ref{fig4c} shows also the cumulative stellar mass function of objects that were identified as having been disrupted at level 4 but which survive at level 2 (set 1; blue dashed). These objects nearly all contain $\lesssim 10^6\, \rm M_{\odot}$ in stellar mass (with the exception of a single $M_* \sim 10^8 \, \rm M_{\odot}$ satellite). Together with the surviving level 4 satellites, these account for approximately one third of the level 2 satellite population; however, the rest, approximately 40 satellites with stellar masses $\lesssim 10^5\, \rm M_{\odot}$, simply do not form at level 4.

We now compare the evolution of the satellites in the standard (level 4) simulation with their high-resolution (level 2) counterparts. In Fig.~\ref{fig5}, we show the evolution of the satellite distance from the centre of the main halo (top row), the total and stellar satellite mass (middle row), and the dark matter density (bottom row) for 3 different satellites, each with a different orbital eccentricity at infall. Firstly, the figure demonstrates that the method described above successfully matches pairs of subhaloes from each simulation; all quantities are almost identical before the second pericentric passage of the satellites, after which time the gravitational influence of the main halo becomes important and clearly gives rise to differences in each case. Thin vertical and horizontal lines indicate the time of first pericentric passage after the satellites cross the virial radius of the main halo, and the pericentre at this time (referred to also as the impact parameter) in all cases. 

For each subhalo shown in Fig.~\ref{fig5}, the total satellite mass reaches a peak at early times ($t_{\rm lookback} \gtrsim 12$ Gyr), and begins to show signs of decline just before and after infall into the main halo and this continues with subsequent pericentric passages. For more circular orbits (illustrated by the objects in the left and right columns of the figure), pericentric passages are many and frequent,  leading to steady mass-loss via tidal stripping of the outer dark matter haloes, whereas the highly eccentric orbit (middle column) experiences fewer, but more sudden episodes of tidal stripping coinciding with its relatively few pericentric passages.  The stellar mass of these objects reaches its peak value later than the dark matter mass owing to the continued formation of stars and because the outer dark matter is stripped before the stars. After the first or second pericentric passage, tidal stripping of stars begins to occur, which leads to a turn-over in satellite stellar mass evolution. 

For each object shown in Fig.~\ref{fig5}, the satellite mass drops rapidly before it disrupts in the level 4 simulation, whereas it continues to decrease smoothly in the level 2 simulation. A typical halo mass at which mass loss deviates toward disruption at standard resolution is $\sim 10^7-10^8$ $\rm M_{\odot}$, which translates to approximately 100 dark matter particles. The reason behind the differences in this late-time evolution between standard and high-resolution may be gleaned from the dark matter density evolution in the lower-panels of Fig.~\ref{fig5}, which illustrates clearly that tidal stripping eventually causes the density of the subhaloes to sharply decrease at standard resolution, whereas it continues to steadily increase at high resolution. Evidently, this dramatic decline reduces the dark matter density of satellites to values below that of the background halo, which ensures their rapid disruption.

\begin{figure}
\includegraphics[width=\columnwidth,trim={0.4cm 1.6cm 0.3cm 1.5cm}, clip]{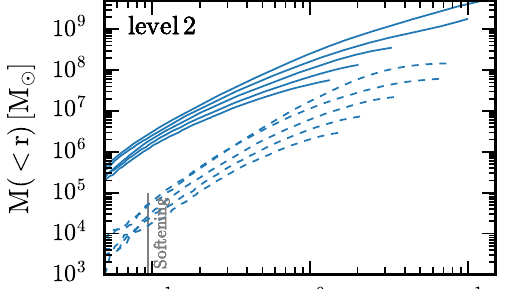}\\
\includegraphics[width=\columnwidth,trim={0.4cm 0.8cm 0.3cm 1.5cm}, clip]{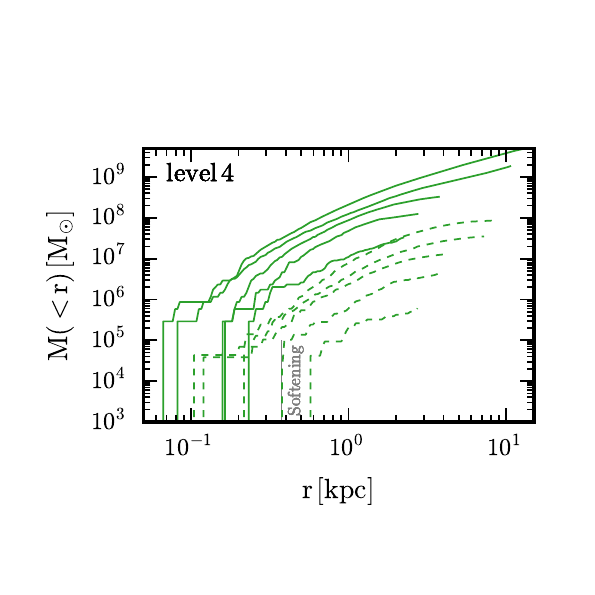}\\
\caption{The cumulative mass radial profile of the dark matter (solid) and stellar matter (dashed) of the satellite presented in the left column of Fig.~\ref{fig5} at each apocentric passage after infall and before the disruption time at standard resolution. The softening length for each simulation is indicated by the vertical line. At the highest resolution, the stripping mainly removes stars from the outer parts of the mass distribution, and leaves the inner parts almost untouched. This is not seen at standard resolution, which instead shows that the inner parts become severely depleted after the third pericentric passage.}
\label{fig6}
\end{figure}

The intriguing difference in dark matter density evolution motivates us to examine the evolution of the cumulative radial mass profile of satellites, which is shown in Fig.~\ref{fig6} for the satellite in the left-column of Fig.~\ref{fig5}. Each curve corresponds to the mass profile measured at an apocentric passage, beginning with the first, and ending with the last apocentric passage of the level 4 satellite. For the satellite shown in Fig.~\ref{fig6}, there are 5 such passages, and profiles are shown at each of these times for dark matter (solid) and stars (dashed). In each case, the mass distribution is reduced mainly from the outside-in from the first to the third apocentric passage, with the half-mass radius for each component decreasing as stripping proceeds. However, after the next apocentric passage, the central parts of the level 4 satellite experience severe mass-depletion in addition to the outer parts, and disrupts soon thereafter. In contrast, the level 2 counterpart satellite largely preserves its central mass content, which is crucial to retain the high density and short crossing timescale that subhaloes require to resist tidal disruption.

\begin{figure*}
\includegraphics[width=\columnwidth,trim={0.3cm 0 0.3cm 0.5cm}, clip]{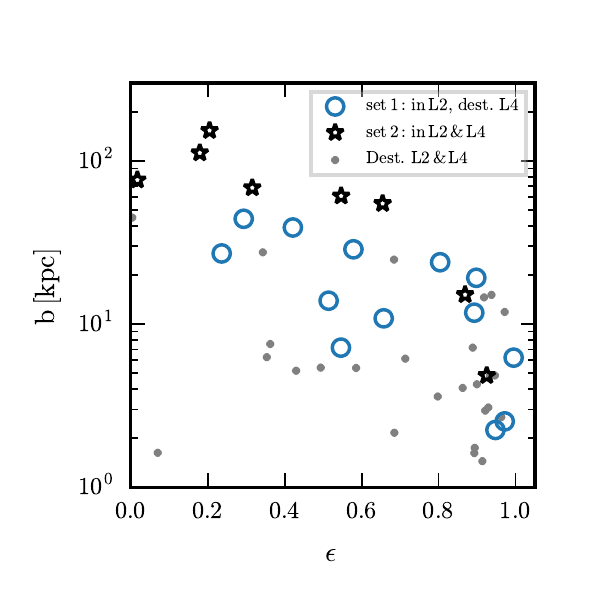}
\includegraphics[width=\columnwidth,trim={0.3cm 0 0.3cm 0.5cm}, clip]{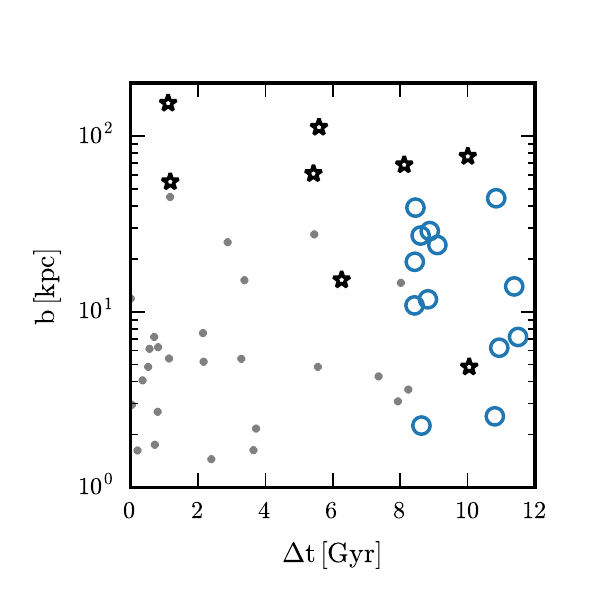}
\caption{Impact parameter (defined as the distance of the first pericentric passage after infall) as a function of orbital eccentricity (defined at infall; left panel), and as a function of the time since infall for surviving satellites or the elapsed time between infall and disruption for disrupted systems (right panel), all measured in the highest resolution simulation. Satellites that survive at both standard and high resolution (stars), satellites that survive only at high resolution (open circles), and satellites that are disrupted in both (solid circles) are shown}
\label{fig7}
\end{figure*}

The efficacy of tidal disruption on subhaloes is expected to depend on properties such as eccentricity and impact parameter; satellites with low impact parameters experience stronger tidal forces than those on larger orbits, and the timescale for disruption is shorter for less eccentric orbits owing to more frequent pericentric passages. This has been demonstrated recently in idealised simulations \citep[e.g.][]{EN21} for systematic studies of tailored situations. We explore the dependencies and implications of our sample of disrupted and surviving satellites at standard resolution and their high-resolution counterparts in Fig.~\ref{fig7}. This figure shows three types of satellites: i) those that survive at both resolution levels; ii) those that survive only at high-resolution; and iii) those that are destroyed at both resolution levels. The left panel shows the impact parameter as a function of eccentricity as measured in the high-resolution level 2 simulation\footnote{Note that the values for the impact parameter, eccentricity and $\Delta t$ (the last of these for surviving satellites only) are almost identical for the standard level 4 resolution simulation, so we plot only the high-resolution level 2 set of measurements for clarity.}. There is a clear trend of lower impact parameters with increasing eccentricity for all surviving satellites (stars and open circles), because stronger tides act faster to disrupt nearer and less-eccentric objects. Interestingly, the satellites that survive only at high resolution populate lower impact parameter-less eccentric regions of this space, i.e., at fixed eccentricity, the impact parameter is smaller compared to satellites that survive at both resolutions. In these regions of parameter space, stronger tides are expected to operate on shorter timescales. Objects disrupted in both simulations (solid dots) are located at even lower impact parameters at fixed eccentricity.

The right-panel of Fig.~\ref{fig7} shows the impact parameter of the same objects plotted now as a function of either the ``infall time'', defined as the lookback time of first infall (for surviving satellites) or ``disruption time'' (for disrupted objects) defined as the time between first infall and disruption. Both are denoted $\Delta t$. It is clear that the satellites that survive independently of resolution have a range of infall times, whereas the objects that survive at high-resolution only populate a region of parameter space characterised by mainly early infall times, which implies satellites that first cross the virial radius at early times are less likely to survive in standard resolution simulations because they experience tidal forces for longer periods of time compared to those with later infall times. Satellites that are destroyed in both simulations show a weak trend for shorter disruption times for lower impact parameters, which again supports expectations. Finally, the majority of satellites that disrupt in level 4 and level 2 do so shortly after infall, as they are mainly on radial orbits and therefore merge rapidly.

\subsubsection{Radial distribution of satellites}
\label{raddist}

Recent observations of satellites of Milky Way-mass host galaxies in the local volume have revealed that, on average, the radial distribution of luminous satellites around their host halo centres is significantly more compact than seen in hydrodynamical simulations \citep{CGP20}, although not in disagreement with semi-analytic simulations \citep{BDB20}. It has been suggested that artificial subhalo disruption in the hydrodynamic simulations may explain this discrepancy as resulting from the destruction of satellites that come close to the central galaxy of their host halo.

\begin{figure}
\includegraphics[width=\columnwidth,trim={0.2cm 0.8cm 0.3cm 1.2cm}, clip]{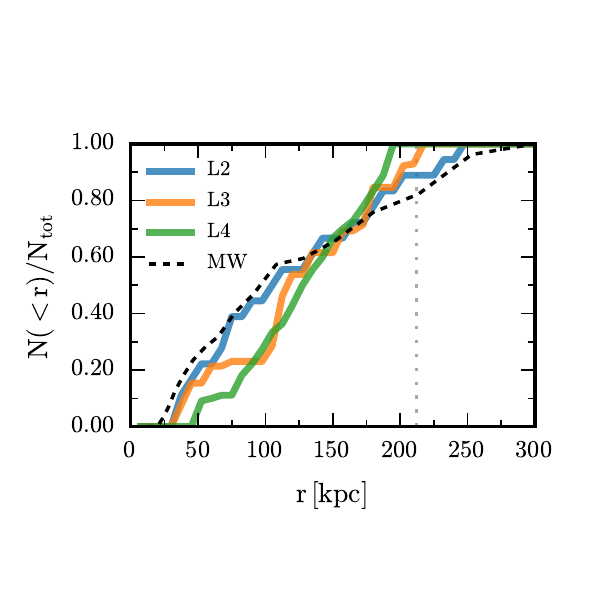}
\caption{The normalised radial distribution of satellites with stellar mass $\geq 5\times 10^4$ $\rm M_{\odot}$ (the stellar mass resolution of our standard level 4 simulation) for the three highest resolution simulations (coloured curves) and for the Milky Way (black dashed curve). The distribution is more centrally concentrated at higher resolution: the innermost tercile is found within 100 kpc at both standard resolution and the level 3 resolution simulations, compared to $\sim70$ kpc at highest resolution. The virial radius is shown by the vertical dotted grey line.}
\label{fig8}
\end{figure}

To explore how the distribution of satellites around their host galaxy centre is affected by resolution, we show in Fig.~\ref{fig8} the radial distribution of satellites normalised by the total number of satellites within the virial radius of the main halo for our 3 highest resolution simulations. We include only those satellites with a stellar mass $M_* \geq 5 \times 10^4 \, \rm M_{\odot}$ in each simulation; this is the stellar mass resolution of our standard level 4 simulation and ensures we compare fairly between our different resolution levels. We note that the radial distribution of satellites can be somewhat time-dependent as they evolve through their orbital phases, and it has been suggested that simulation snapshots should be time-averaged to mitigate this dependence \citep{Sawala_17,SWT20}. We therefore time-average the radial distribution over the last several snapshots of each simulation (approximately the last gigayear of evolution). This is comparable to the orbital timescale of a typical satellite in the inner halo, and therefore should be sufficient to mitigate the time-dependence of the radial distribution in the main region of interest for our analysis.

Figure~\ref{fig8} shows a clear trend in which the distribution becomes more centrally concentrated as numerical resolution increases: the inner tercile of satellite distances extends to $\sim 70$ kpc in the level 2 simulation, compared to $\sim 100$ kpc in both the level 4 and level 3 simulations. We note that if we include only objects with stellar masses $> 10^6$ $\rm M_{\odot}$, the radial distribution is similar at all resolution levels, which indicates that these objects are well-resolved at standard resolution (although note the single level 2 object of $\sim 10^8$ $\rm M_{\odot}$ stellar mass that disrupts at standard resolution, Fig.~\ref{fig5}). However, we caution that for the system presented in this paper, there are very few objects at these masses.

Alongside the simulation curves in Fig.~\ref{fig8}, we show also the completeness-corrected distribution of Milky Way satellites derived by \citet{NCJ18} (see Fig. 3 of that study), which compensates for the radial bias in the raw observations. Although these data represent only one host galaxy, the level of agreement of the observations with our level 2 simulation, particularly within a galactocentric radius of $\sim 150$ kpc, is striking.

To extend our comparison to galaxies in the local volume \citep[in particular to those presented in][]{CGP20}, we construct synthetic projected radial distributions of our simulated satellites to mimic the selection function of the observations. To this end, we first select satellites with $M_V < -12$ and $\mu _{V,\rm eff} < 28.3$ (as described in section~\ref{subpop}). To increase the sample size and statistical significance of our results, for each simulation, we: i) time average over the same period mentioned above; and ii) calculate the projected radius of satellites along 1000 random lines of sight for each snapshot. Further, for the standard resolution, we average over the 7 different realizations that we have available. 

\begin{figure}
\includegraphics[width=\columnwidth,trim={0.2cm 0.8cm 0.3cm 1.2cm}, clip]{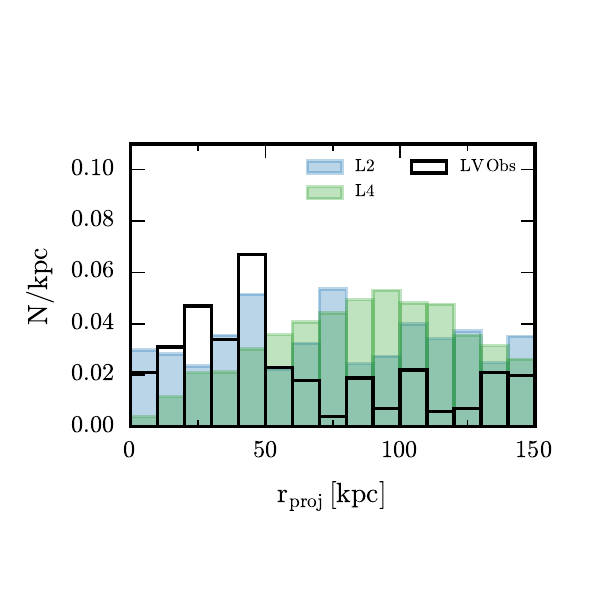}
\caption{Histogram of the projected radial distance of mock samples of satellites for our simulated standard (green) and high (blue) resolution simulations (see text for details). These histograms are normalised in a similar way to the local volume observations (black) presented in \citet{CGP20}. The projected radial distribution of satellites in the high resolution simulation has median and lower quartile values equal to 75 kpc and 40 kpc, respectively, compared to values of 90 kpc and 60 kpc for the standard simulations. The former set of values agrees better with the observations than do the latter.}
\label{fig8b}
\end{figure}

Figure~\ref{fig8b} shows the histogram of the projected radial distance of the synthetic satellite sample at the standard resolution and high resolution out to 150 kpc, which is approximately the aperture used in the observations \citep{CGP20}. It is clear that the projected radial distribution of simulated satellites is more centrally concentrated compared to the standard resolution: the median and lower quartile projected distances for the high resolution simulation are 75 kpc and 40 kpc, respectively, compared to values of 90 kpc and 60 kpc for the standard simulations. The high resolution simulation is thus a better match to the local volume data. We may infer from the results shown in Figs~\ref{fig8} and \ref{fig8b} that the limited  resolution of typical cosmological hydrodynamics simulations causes their distribution of satellites with $M_* \gtrsim 10^5 \, \rm M_{\odot}$ to be less concentrated than certainly higher resolution simulations, and perhaps even observations, which supports the recent findings of semi-analytic models applied to high-resolution N-body simulations \citep{BDB20}. Note that, aside from resolution, other possible explanations for the apparent discrepancy between predicted and observed distributions of satellites have been put forward (see Sec.~\ref{Discussion} below).

\subsubsection{Satellite scaling relations}

In this section, we demonstrate that our simulation is not only capable of following the formation of the full luminosity range of Milky Way satellites (including ultra-faint galaxies), but also that these simulated satellites match a range of observational scaling relations across $\sim 6$ orders of magnitude in $V$-band luminosity. 

\begin{figure}
\includegraphics[width=\columnwidth,trim={0.3cm 1.6cm 0.4cm 1.cm}, clip]{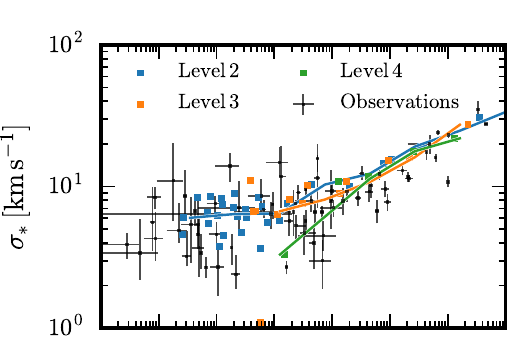}\\
\includegraphics[width=\columnwidth,trim={0.3cm 0.5cm 0.4cm 1.3cm}, clip]{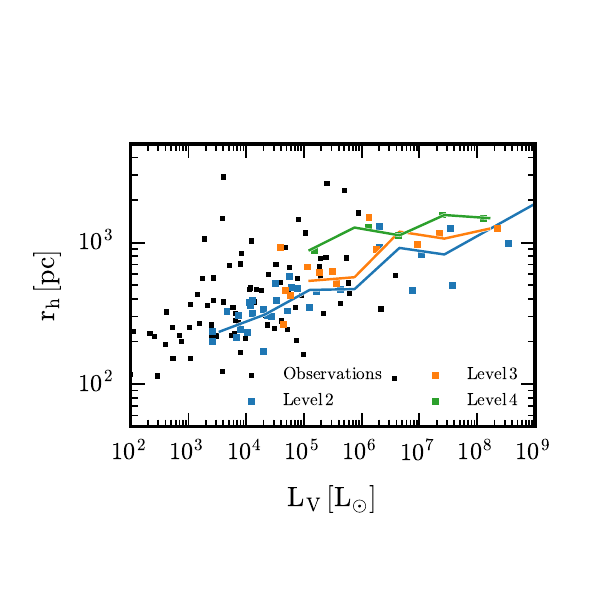}
\caption{Top: stellar velocity dispersion as a function of $V$-band luminosity for satellite galaxies within 1 Mpc of the central galaxy in our three highest resolution simulations. Squares represent individual satellites, and the median velocity dispersion for each luminosity is shown by the solid curves. Bottom: as the top panel, but for the stellar $V$-band half-light radii of satellite galaxies. We include only satellites that contain at least 10 star particles. Observations for Milky Way and M31 satellites are shown in black \citep{McConnachie12} (following \citet{ABC21}, in the lower panel we exclude satellites with half-light radii less than 100 pc, because this is approximately our level 2 softening length and also the regime where it is difficult to distinguish dwarf galaxies from globular clusters). The highest resolution simulation probes well into the ultra-faint galaxy regime, and agrees well with both observed  scaling relations.}
\label{fig9}
\end{figure}

The top panel of Fig.~\ref{fig9} shows the stellar velocity dispersion of satellite galaxies within 300 kpc of the main halo centre for our three highest resolution simulations as a function of $V$-band luminosity (relative to Solar). The median values of stellar velocity dispersion in bins of luminosity are depicted by the solid curves. At the bright end ($L_V \gtrsim 10^7 \, \rm L_{\odot}$), all simulations show almost identical relations that agree well with observations. Proceeding from brighter to fainter luminosities, the standard resolution simulation curve drops below the observational data points for luminosities fainter than $L_V \sim 10^6 \, \rm L_{\odot}$, before cutting off entirely a further order of magnitude lower. This trend is replicated for the next highest resolution simulation (level 3), but shifted an order of magnitude fainter, whereas our highest resolution simulation probes satellites as faint as $L_V \sim 10^3 \, \rm L_{\odot}$ - well into the ultra-faint galaxy regime. We note that for this simulation, the median appears fully consistent with the observed Milky Way satellites over the entire luminosity range. We note that the scatter at $L_V \sim 10^6 \, \rm L_{\odot}$ appears larger in the observations compared to the simulation, but this may be because the observations contain objects from more than one system whereas we have only one simulated galaxy with few objects at this luminosity.

The lower panel of Fig.~\ref{fig9} shows the half-light radius of satellites within 300 kpc of main halo centre for our three highest resolution simulations as a function of $V$-band luminosity. The same trends seen in the upper panel of Fig.~\ref{fig9} hold for this relation, except that the half-light radius decreases for the magnitude range $L_V \sim 10^4 \, \rm L_{\odot}$ - $10^6 \, \rm L_{\odot}$ with increasing resolution, which brings the level 2 resolution simulation into good agreement with the observations. We note that there are a few observed large, faint satellites not found among the simulated satellites. Whether this is because of a lack of (sub)halo diversity inherent to a single simulated galaxy, a manifestation of our galaxy formation model, or of numerical origin, is unclear. Nevertheless, that the stellar velocity dispersion and half-light radii of our highest-resolution satellites agree well with the general trends seen in the observations indicates that the Auriga galaxy formation model produces realistic mass distributions for both dark matter and stars in satellite galaxies spanning almost the entire luminosity range of observed Milky Way satellites. 

\begin{figure}
\includegraphics[width=\columnwidth,trim={0cm 0.cm 0.4cm 0.4cm}, clip]{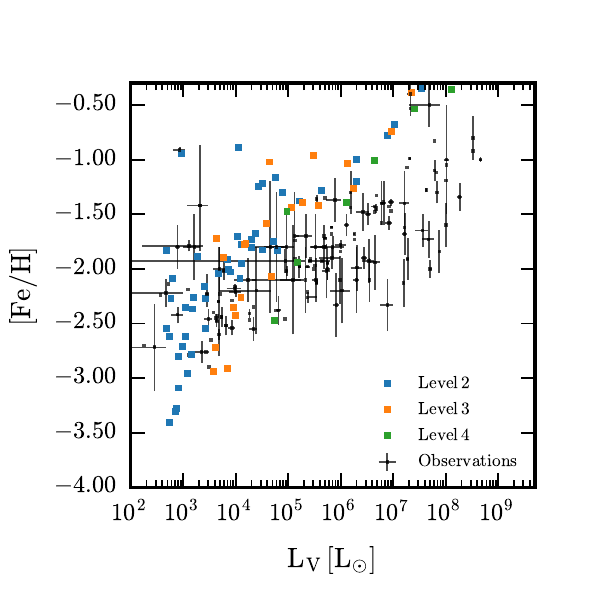}
\caption{Stellar iron abundance of satellites as a function of their $V$-band luminosity for the three highest resolution simulations. Here, we include all satellites with at least 1 star particle within 1 Mpc of the central galaxy. Observations of Milky Way and M31 satellites are shown in black, and are taken from \citet{McConnachie12}, \citet{KCG13}, and \citet{Simon19}.}
\label{fig10}
\end{figure}

In Fig.~\ref{fig10}, we show the metallicity of satellite galaxies as a function of $V$-band luminosity for our three highest resolution simulations. Metallicity increases with brightness as observed, but for magnitudes brighter than $L_V\sim 10^4 \, \rm L_{\odot}$, the metallicity of the simulated objects is systematically $\sim 0.5$~dex too large, although many lie within the observed scatter. Preliminary investigations into the feedback model suggest that this discrepancy can be removed if metal-poor systems are assumed to produce faster supernova-driven winds than metal-rich systems where radiative losses are plausibly more significant \citep[see][]{PSN18a,VGS21}. This scaling increases the efficacy with which winds expel metals at early times, and is expected because additional cooling losses should slow down winds. Nevertheless, for luminosities fainter than $L_V\sim 10^4 \, \rm L_{\odot}$ nearly all of our satellites are consistent with the observed scatter in this relation.

\section{Discussion}
\label{Discussion}

Our study highlights the importance of high resolution for predicting the satellite luminosity function down to the ultra-faint regime. The main reason is that at standard (level 4) resolution most of these objects do not form. This result supports the recent semi-analytic work of \citet{NCJ18}, who applied the Galform model to the high-resolution dark matter-only Aquarius haloes \citep{SWV08}, and showed that accounting for so-called ``orphan'' galaxies and the destruction of satellites by a central disc produced an excellent match to the Milky Way's satellite luminosity function. Our high resolution simulation confirms this in a cosmological hydrodynamic simulation. 

A secondary effect is the disruption of subhaloes as a result of gravitational tides induced by the central dark halo and stellar disc \citep{DSH10,YS15,RFJ20}; approximately one fifth of the satellites in our highest resolution simulation were formed and disrupted before $z=0$ at standard resolution. This indicates that our high resolution simulation at least partially mitigates some artificial disruption effects that tailored $N$-body simulations have highlighted \citep[e.g.][]{BOH18,EN21}. These effects are a concern primarily in the ultra-faint regime, although they seem to play a subdominant role in the ability of the simulations to reproduce the faint-end of the luminosity function compared to the effect of forming faint galaxies in the first place. We note that these effects are ameliorated but remain significant for simulations with a higher-than-standard mass resolution of $\sim 6\times 10^3 \, \rm M_{\odot}$ (equivalent to our level 3 simulation). At face value, this means that even the highest resolution modern cosmological zoom-in simulations may be missing a significant fraction of satellites at the present day. We defer a detailed investigation into these numerical effects to a future study.

In Sec.~\ref{raddist}, we showed evidence that the distribution of satellites around their central host galaxy becomes more centrally concentrated for higher resolution. This result is in agreement with those of \citet{BDB20}, who used the Galform semi-analytic model applied to the COCO $N$-body simulations \citep{BHF16,HFC16} to show that $\Lambda$CDM is consistent with the observed radial distribution of Milky Way satellites after effectively increasing the resolution of their simulations by taking into account orphan galaxies. These results imply that the discrepancies between the radial distribution of satellites in large-box simulations and observations of nearby spiral galaxies discussed by \citet{CGP20} can be explained if sufficiently high resolution is employed. It is also important to note that \citet{FMB21} showed that modelling observational selection effects improves the level of agreement between predictions from the Artemis simulations \citep{FMP20} and Milky Way analogues sampled in the local volume and SAGA surveys \citep{MGW21}. However, the low number of objects in the former survey in particular will require more observations to ascertain the statistical robustness of this result.

Finally, we note that in addition to numerical resolution, differences in the stellar disc properties between our standard and highest resolution simulations may further affect the rates of subhalo disruption. Although less than $2\%$ more massive, the standard resolution stellar disc is 1~kpc larger in scale-length, which may make it more effective at disrupting subhaloes. Furthermore, we caution that our results are derived from a resolution study of a single Milky Way-mass system simulated with a particular galaxy formation model, and that a larger sample of simulations is desirable to have a more statistically robust handle on results we have presented. 

\section{Conclusions}
\label{Conclusions}

We have presented a magneto-hydrodynamic cosmological simulation of the formation of a Milky Way-mass galaxy based on the \textlcsc{AURIGA} model with an unprecedented 800 $\rm M_{\odot}$ mass resolution per baryonic element. We verified that this simulation produces a realistic star-forming spiral disc galaxy, with a radial stellar surface density profile that reflects a disc-dominated system, and a star formation history that peaks at $z\sim1$ and steadily decreases to present-day values of order $1 \, \rm M_{\odot} \, yr^{-1}$. We showed that these quantities and the properties and abundance of subhaloes converge well over 3.5 orders of magnitude in mass resolution, and are thus, for the first time, numerically robust over a large range of scales. We further showed that the Auriga model is also robust to stochastic variations associated with the ``Butterfly Effect'', which are small. 

We analysed the properties of the simulated subhaloes and satellites, and how they depend on numerical resolution. We come to the following main conclusions:

\begin{itemize}
\item{} We show that our highest resolution simulation provides an excellent match to the observed abundance of Milky Way satellites, particularly at the faint end ($M_* \lesssim 10^5 \, \rm M_{\odot}$, $M_V \gtrsim -10$), whereas lower resolution simulations underestimate it significantly. We show that this is mainly because two thirds of high resolution satellites never form at standard resolution, and that a smaller number of satellites that do form in both simulations are disrupted at standard resolution before $z=0$.
\item{} We find that the radial distribution of satellites becomes more compact with increasing resolution: the median normalised satellite distance shifts from $\sim 130$ kpc at standard resolution to $\sim 100$ kpc at the highest resolution, in better agreement with the radial distribution of Milky Way satellites. This trend disappears if only satellites more massive than $\sim 10^6 \, \rm M_{\odot}$ in stars are considered, reflecting the effects of high resolution faint satellites that either do not form or are disrupted at lower resolution. Among the latter type, objects on low-eccentricity and low-impact parameter orbits are particularly affected. Further, we show that mock observations of the projected radii of high-resolution satellites reproduce the observed trend in local volume observations, in contrast to standard resolution satellites whose radial distribution is much too extended.
\item{} We show that our highest resolution simulation reproduces the mean stellar velocity dispersion-luminosity and half-light radius-luminosity scaling relations of Milky Way and Local Group satellites in the luminosity range $L_V \sim 10^3 \, \rm L_{\odot}$ - $10^{10} \, \rm L_{\odot}$. However, the stellar metallicities of satellites more luminous than $L_V \sim 10^5 \, \rm L_{\odot}$ are higher than observed by about 0.5 dex. More simulations are required to attain the statistics necessary to determine whether the scatter in the observed relations is fully reproduced.
\end{itemize}

Highly resolved cosmological hydrodynamic simulations such as the one presented in this paper provide exquisite sampling of the phase space of galactic stellar structure that could be compared to deep photometric observations of external galaxies (e.g. SDSS), and Galactic surveys such as {\it Gaia} and upcoming spectroscopic surveys (e.g. 4MOST) that are regularly uncovering new and more detailed Galactic (sub)structure. Moreover, such a highly resolved central galaxy ($\gtrsim 10^8$ star particles in the disc) surpasses a threshold identified in idealised $N$-body simulations required to effectively eliminate Poisson noise as a seed for the formation of spiral arms \citep{DO12}. Thus, simulations of this type should be useful for investigations related to galactic dynamics and archaeology.

\section*{Acknowledgements}
We acknowledge the referee for a thoughtful and helpful report. RG thanks Oliver Newton for sharing his data on the radial distribution of satellites in the Milky Way, and Marius Cautun for helpful suggestions. RG acknowledges financial support from the Spanish Ministry of Science and Innovation (MICINN) through the Spanish State Research Agency, under the Severo Ochoa Program 2020-2023 (CEX2019-000920-S). CSF acknowledges support from the European Research Council through ERC Advanced Investigator grant, DMIDAS [GA 786910]. This work was also supported by STFC Consolidated Grants for Astronomy at Durham ST/P000541/1 and ST/T000244/1. It used the DiRAC Data Centric system at Durham University, operated by the Institute for Computational Cosmology on behalf of the STFC DiRAC HPC Facility (www.dirac.ac.uk). This equipment was funded by BIS National E-infrastructure capital grants ST/P002293/1,
ST/R002371/1 and ST/S002502/1, Durham University and STFC operations grant ST/R000832/1. FAG acknowledges financial support from CONICYT through the project FONDECYT Regular Nr. 1211370, and funding from the Max Planck Society through a Partner Group grant. FM acknowledges support through the program ``Rita Levi Montalcini'' of the Italian MIUR. AJK acknowledges an STFC studentship grant ST/S505365/1.

\section*{Data Availability}
The data underlying this article will be shared on reasonable request to the corresponding author.

\bibliographystyle{mnras}
\bibliography{pap1.bib}
%\bibliography{mnras_template.bbl}

% Don't change these lines
\bsp	% typesetting comment
\label{lastpage}
\end{document}